\documentclass[aps,prx,twocolumn,tightenlines,showpacs,superscriptaddress]{revtex4-1} %% Ben's package
\usepackage[utf8]{inputenc}
\usepackage[T1]{fontenc}
\usepackage{amsfonts}
\usepackage{dcolumn}
\usepackage{bm}
\usepackage{amsmath}
\usepackage{nicefrac}
\usepackage{amsthm}
\usepackage{amssymb}
\usepackage{graphicx}
\usepackage{color}
\usepackage{url}
\usepackage{cancel}
\usepackage{units}
\usepackage{upgreek}
\usepackage[shortcuts]{extdash}
\usepackage[normalem]{ulem}

\usepackage{hyperref} % Required for adding links	and customizing them
\definecolor{linkcolour}{rgb}{0.,0.,0.8} % Link color
\hypersetup{colorlinks=true,urlcolor=linkcolour,linkcolor=linkcolour,citecolor=linkcolour} % Set link colors throughout the document

\newcommand{\appropto}{\mathrel{\vcenter{
  \offinterlineskip\halign{\hfil$##$\cr
    \propto\cr\noalign{\kern2pt}\sim\cr\noalign{\kern-2pt}}}}}
\renewcommand{\v}[1]{\boldsymbol{#1}}		%bold-math for vectors

\newcommand{\note}[1]{{\color{blue}[#1]}}
\renewcommand{\note}[1]{}

\begin{document}

%%%%%%%%%%%%%%%%%%%%%%%%%%%%%%%%%%%%%%%%%%%%%%%%%%
%\title{Search for Axion-like Dark Matter through the Oscillation of the Neutron Electric Dipole Moment {\color{black} and Axion-Wind Spin-Precession Effect}}
\title{Search for axion-like dark matter through nuclear spin precession in electric and magnetic fields}

\date{\today}

% \author{N.~J.~Ayres$^{1}$} %\email[]{}
% \author{M.~Fairbairn$^{2}$} %\email[]{}
% \author{V.~V.~Flambaum$^{3,4}$} %\email[]{}  %% Add Helmholtz Institut of Mainz affiliation also?
% \author{P.~G.~Harris$^{1}$} %\email[]{}
% \author{K.~Kirch$^{5,6}$} %\email[]{}
% \author{D.~J.~E.~Marsh$^{2}$} %\email[]{}
% \author{G.~Pignol$^{7}$} %\email[]{}
% \author{M.~Rawlik$^{6}$} %\email[]{}
% \author{Y.~V.~Stadnik$^{3,4}$} %\email[]{}

\author{C.~Abel}
\affiliation{Department of Physics and Astronomy, University of Sussex, Falmer, Brighton BN1 9QH, United Kingdom}

\author{N.\,J.~Ayres}
\email[Corresponding author,\ ]{N.Ayres@sussex.ac.uk}
\affiliation{Department of Physics and Astronomy, University of Sussex, Falmer, Brighton BN1 9QH, United Kingdom}

% \author{\color{red} C.\,A.~Baker}
% \affiliation{Rutherford Appleton Laboratory, Chilton, Didcot, Oxon OX11 0QX, UK}
% no response

\author{G.~Ban}
% \affiliation{Normandie Univ, ENSICAEN, UNICAEN, CNRS/IN2P3, LPC Caen, 14000 Caen, France}
% gave no response

\author{G.~Bison}
\affiliation{Paul Scherrer Institute, CH-5232 Villigen PSI, Switzerland}

\author{K.~Bodek}
\affiliation{Institute of Physics, Jagiellonian University in Krakow, Poland}
% wants to be an author
% didn't comment on the paper

\author{V.~Bondar}
\affiliation{Instituut voor Kern en Stralingsfysica, Katholieke Universiteit Leuven, B-3001 Leuven, Belgium}

\author{M.~Daum}
\affiliation{Paul Scherrer Institute, CH-5232 Villigen PSI, Switzerland}

\author{M.~Fairbairn} %\email[]{}
\affiliation{King's College London Department of Physics, London, WC2R 2LS, United Kingdom}

\author{V.\,V.~Flambaum} %\email[]{}
\affiliation{School of Physics, University of New South Wales, Sydney 2052, Australia}
\affiliation{Johannes Gutenberg University, 55122 Mainz, Germany}

\author{P.~Geltenbort}
\affiliation{Institut Laue-Langevin, BP 156, F-38042 Grenoble Cedex 9, France}
% no response

\author{K.~Green}
\affiliation{Rutherford Appleton Laboratory, Chilton, Didcot, Oxon OX11 0QX, UK}
% no response

\author{W.\,C.~Griffith}
\affiliation{Department of Physics and Astronomy, University of Sussex, Falmer, Brighton BN1 9QH, United Kingdom}

\author{M.~van~der~Grinten}
\affiliation{Rutherford Appleton Laboratory, Chilton, Didcot, Oxon OX11 0QX, UK}
% no response

\author{Z.\,D.~Gruji\'c}
\affiliation{Physics Department, University of Fribourg, Fribourg, Switzerland}

\author{P.\,G.~Harris} %\email[]{}
\affiliation{Department of Physics and Astronomy, University of Sussex, Falmer, Brighton BN1 9QH, United Kingdom}

\author{N.~Hild}
\affiliation{Paul Scherrer Institute, CH-5232 Villigen PSI, Switzerland}

\author{P.~Iaydjiev}
\affiliation{Rutherford Appleton Laboratory, Chilton, Didcot, Oxon OX11 0QX, UK}
\affiliation{Present address: Institute of Nuclear Research and Nuclear Energy, Sofia, Bulgaria}
% no response

\author{S.\,N.~Ivanov}
\affiliation{Rutherford Appleton Laboratory, Chilton, Didcot, Oxon OX11 0QX, UK}
\affiliation{Present address: Petersburg Nuclear Physics Institute, Russia}
% no response

\author{M.~Kasprzak}
\affiliation{Instituut voor Kern en Stralingsfysica, Katholieke Universiteit Leuven, B-3001 Leuven, Belgium}

\author{Y.~Kermaidic}
\affiliation{LPSC, Universit{\'e} Grenoble Alpes, CNRS/IN2P3, Grenoble, France}

\author{K.~Kirch}
\affiliation{ETH Z{\"u}rich, Institute for Particle Physics, CH-8093 Z{\"u}rich, Switzerland}
\affiliation{Paul Scherrer Institute, CH-5232 Villigen PSI, Switzerland}

\author{H.-C.~Koch}
\affiliation{Paul Scherrer Institute, CH-5232 Villigen PSI, Switzerland}

\author{S.~Komposch}
\affiliation{Paul Scherrer Institute, CH-5232 Villigen PSI, Switzerland}
\affiliation{ETH Z{\"u}rich, Institute for Particle Physics, CH-8093 Z{\"u}rich, Switzerland}

\author{P.\,A.~Koss}
\affiliation{Instituut voor Kern en Stralingsfysica, Katholieke Universiteit Leuven, B-3001 Leuven, Belgium}

\author{A.~Kozela}
\affiliation{Institute of Nuclear Physics, Polish Academy of Sciences, Krakow, Poland}

\author{J.~Krempel}
\affiliation{ETH Z{\"u}rich, Institute for Particle Physics, CH-8093 Z{\"u}rich, Switzerland}

\author{B.~Lauss}
\affiliation{Paul Scherrer Institute, CH-5232 Villigen PSI, Switzerland}

\author{T.~Lefort}
\affiliation{Normandie Univ, ENSICAEN, UNICAEN, CNRS/IN2P3, LPC Caen, 14000 Caen, France}

\author{Y.~Lemière}
\affiliation{Normandie Univ, ENSICAEN, UNICAEN, CNRS/IN2P3, LPC Caen, 14000 Caen, France}

\author{D.\,J.\,E.~Marsh} %\email[]{}
\affiliation{King's College London Department of Physics, London, WC2R 2LS, United Kingdom}

\author{P.~Mohanmurthy}
\affiliation{Paul Scherrer Institute, CH-5232 Villigen PSI, Switzerland}
\affiliation{ETH Z{\"u}rich, Institute for Particle Physics, CH-8093 Z{\"u}rich, Switzerland}

\author{A.~Mtchedlishvili}
\affiliation{Paul Scherrer Institute, CH-5232 Villigen PSI, Switzerland}

\author{M.~Musgrave}
\affiliation{Laboratory for Nuclear Science, Massachusetts Institute of Technology, Cambridge, MA 02139, USA}

\author{F.\,M.~Piegsa}
\affiliation{Laboratory for High Energy Physics, University of Bern, CH-3012 Bern, Switzerland}

\author{G.~Pignol} %\email[]{}
\affiliation{LPSC, Universit{\'e} Grenoble Alpes, CNRS/IN2P3, Grenoble, France}

\author{M.~Rawlik}
\email[Corresponding author,\ ]{mrawlik@phys.ethz.ch}
\affiliation{ETH Z{\"u}rich, Institute for Particle Physics, CH-8093 Z{\"u}rich, Switzerland}

\author{D.~Rebreyend}
\affiliation{LPSC, Universit{\'e} Grenoble Alpes, CNRS/IN2P3, Grenoble, France}

\author{D.~Ries}
\affiliation{Laboratory for High Energy Physics, University of Bern, CH-3012 Bern, Switzerland}
\affiliation{Paul Scherrer Institute, CH-5232 Villigen PSI, Switzerland}
\affiliation{ETH Z{\"u}rich, Institute for Particle Physics, CH-8093 Z{\"u}rich, Switzerland}

\author{S.~Roccia}
\affiliation{CSNSM, Université Paris Sud, CNRS/IN2P3, Orsay campus, France}

\author{D.~Rozpędzik}
\affiliation{Institute of Physics, Jagiellonian University in Krakow, Poland}

\author{P.~Schmidt-Wellenburg}
\affiliation{Paul Scherrer Institute, CH-5232 Villigen PSI, Switzerland}

\author{N.~Severijns}
\affiliation{Instituut voor Kern en Stralingsfysica, Katholieke Universiteit Leuven, B-3001 Leuven, Belgium}

\author{D.~Shiers}
\affiliation{Department of Physics and Astronomy, University of Sussex, Falmer, Brighton BN1 9QH, United Kingdom}

\author{Y.\,V.~Stadnik} %\email[]{}
\affiliation{School of Physics, University of New South Wales, Sydney 2052, Australia}
\affiliation{Johannes Gutenberg University, 55122 Mainz, Germany}

\author{A.~Weis}
\affiliation{Physics Department, University of Fribourg, Fribourg, Switzerland}

\author{E.~Wursten}
\affiliation{Instituut voor Kern en Stralingsfysica, Katholieke Universiteit Leuven, B-3001 Leuven, Belgium}

\author{J.~Zejma}
\affiliation{Institute of Physics, Jagiellonian University in Krakow, Poland}

\author{G.~Zsigmond}
\affiliation{Paul Scherrer Institute, CH-5232 Villigen PSI, Switzerland}

\begin{abstract}
We report on a search for ultra-low-mass axion-like dark matter by analysing the ratio of the spin-precession frequencies of stored ultracold neutrons and $^{199}$Hg atoms for an axion-induced oscillating electric dipole moment of the neutron {\color{black}and an axion-wind spin-precession effect}. %and coupling of an oscillating galactic axion field to the nucleon spins.
No signal consistent with dark matter is observed for the axion mass range $10^{-24}~\textrm{eV} \le m_a \le 10^{-17}~\textrm{eV}$.
Our null result sets the first laboratory constraints on the coupling of axion dark matter to gluons, which improve on astrophysical limits by up to 3 orders of magnitude, {\color{black} and also improves on previous laboratory constraints on the axion coupling to nucleons by up to a factor of 40}.
%\note{CG: not obvious to me in Fig.\,4 that the new lines are 3 orders of magnitude below the red region (in the non excluded part to the right of the orange region).}
%\note{YS: Fig.~4 assumes that axions saturate the observed DM content. At the threshold of the orange limit at $m=10^{-22}$ eV, our results in blue improve over the astrophysical bounds in red by 2 orders of magnitude. On the other hand, if axions make up only a sub-dominant fraction of the total DM, then the constraints in orange go very far to the left (out of our reach), while the blue nEDM and red BBN constraints are both shifted upwards by the same amount, so in this case there is up to 3 orders of magnitude improvement. Any suggestions on how to word this economically in the text/figure would be great.}
%[To add later: First such search for axions with UCNs; result is interpretted as the limits (maximum sensitivities for which masses) on the interactions x, y, and z, which improve on existing astrophysical/laboratory bounds for the masses (ranges)...]
\end{abstract}

%\pacs{95.35.+d, 14.80.Va, 14.20.Dh, 11.30.Er}    %% Dark Matter, Axions, Nucleons, Fundamental Symmetries

\maketitle

%%%%%%%%%%
\section{Introduction}
\label{Sec:Introduction}
% \textbf{Introduction.} ---
%\note{Text formatted like this are notes and comments, not part of the final text.}

%\note{GP, TL, DR: Earlier searches for oscillating signals using our apparatus should be mentioned and cited.
%namely:
%24h modulation of R: I. Altarev et al Phys.Rev.Lett. 103 (2009) 081602
%12h and 24h modulation of nEDM: I. Altarev et al. Europhys.Lett. 92 (2010) 51001}
% \note{KK: Comment explicitly on the phase of the signal being random in the simulations. We are not looking at the worst case scenario!}
%\note{MMM: I would recommend subdividing your paper into labeled sections for the benefit of the reader.}
%\note{MR: PRL papers have no sections, right?}
%\note{YS: PRL letters are not allowed to have the usual sections (like in the other subdivisions of Physical Review), but it is possible to have small subheadings throughout the text, if this would help:~e.g., \textbf{Introduction.} --- Text ...}

%\note{YS: In the list of references, footnote [18] appears to have been duplicated in [66]. Is it possible to have the footnote appear only once in the references (as [18]), like we had before?}
%note{MR: Note. \footnotemark does the trick.}

% \note{KK: Remove the title from Ref. [15].}

% \note{YS: In Ref.~[81], remove ''[On-line; accessed date]'', and replace by url.}

Astrophysical and cosmological observations indicate that $26 \%$ of the total energy density and $84 \%$ of the total matter content of the Universe is dark matter (DM) \cite{Planck2015}, %\note{FP: Right at the beginning you say there is 85\% dark matter - normally its stated that there is 27\% DM and 68\% Dark Energy - can you just add this up to one number?} \note{PSW: I wouldn't sum over dark matter and dark energy. Both have different observational indicators and interpretations.} \note{Doddy: Reword to "26\% of the total energy density of the Universe, and 84\% of the total matter content"} \note{YS: I agree with Doddy's suggestion.},
the identity and properties of which still remain a mystery.
One of the leading candidates for cold DM
%\note{PMM: \ldots candidate for cold dark matter. cite: Phys. Rev. Lett. 108, 061304 also. Axion being hot dark matter needs more exotic interactions.}
%\note{YS: The general consensus in cosmology is that the vast majority of DM should be cold, so the term DM without qualifiers is often taken to be synonymous with cold DM. Hot DM is quite exotic (the axion is no exception). On a side note, the mentioned paper refers to the formation of an axion BEC with an enormous correlation length (much larger than the standard assumption of the de Broglie wavelength associated with the virialised component of the axions) - this is quite a contraversial claim, which is disputed by a sizeable portion of the physics community.}
is the axion, a pseudoscalar particle which was originally hypothesised to resolve the strong CP problem of quantum chromodynamics (QCD) \cite{PQ1977A,PQ1977B,Weinberg1978,Wilczek1978,Kim1979,Zakharov1980,Zhitnitsky1980A,*Zhitnitsky1980B,Srednicki1981}.
Apart from the canonical QCD axion, various axion-like particles have also been proposed, for example in string compactification models \cite{Witten1984,Conlon2006,Witten2006,Arvanitaki2010,Arias2012,Marsh2015Review}.

Low-mass ($m_a \lesssim 0.1~\textrm{eV}/c^2$) axions can be produced efficiently via non-thermal production mechanisms, such as vacuum misalignment \cite{Preskill1983cosmo,Sikivie1983cosmo,Dine1983cosmo} in the early Universe, and subsequently form a coherently oscillating classical field \footnote{Although non-thermal production mechanisms typically impart negligible kinetic energy to the produced axions, the gravitational interactions between axions and ordinary matter during galactic structure formation subsequently virialise galactic axions ($v_{\textrm{vir}}^{\textrm{local}} \sim 300$~km/s), giving an oscillating galactic axion field the finite coherence time:~$\tau_{\textrm{coh}} \sim 2\pi / m_a v_{\textrm{vir}}^2 \sim 2\pi \cdot 10^6 / m_a $, i.e., $\Delta \omega / \omega \sim 10^{-6}$. }:~$a = a_0 \cos(\omega t)$, with the angular frequency of oscillation given by $\omega \approx m_a c^2 / \hbar$, where $m_a$ is the axion mass (henceforth, we shall adopt the units $\hbar = c = 1$).
%\note{PMM: replace with ...classical field - a, $a=a_0 cos(wt)$, with the angular frequency of oscillations given by $\omega \sim m_a$, where $m_a$ is the axion mass.
%No need to mention $c=\hbar = 1$ as you've already used natural units before. Maybe you can mention in brackets: ... $\omega \sim m_a$ (in natural units). This is the first time $m_a$ is used so needs to be explicitly mentioned as the axion mass.}
The oscillating axion field carries the energy density $\rho_a \approx m_a^2 a_0^2 /2$. Due to its effects on structure formation \cite{Khlopov1985}, ultra-low-mass axion DM in the mass range $10^{-24}~\textrm{eV} \lesssim m_a \lesssim 10^{-20}$ eV has been proposed as a DM candidate that is observationally distinct from, and possibly favourable to, archetypal cold DM \cite{Hu2000,Marsh2014,Schive2014,Marsh2015Review,Hui2017}.
The requirement that the axion de Broglie wavelength does not exceed the DM size of the smallest dwarf galaxies and consistency with observed structure formation \cite{Marsh2015B,Schive2015,Marsh2017} give the lower axion mass bound $m_a \gtrsim 10^{-22}$ eV, if axions comprise all of the DM. However, axions with smaller masses can constitute a sub-dominant fraction of DM~\cite{Hlozek15}.

It is reasonable to expect that axions interact non-gravitationally with standard-model particles.
Direct searches for axions have thus far focused mainly on their coupling to the photon (see the review \cite{Axion-Review2015} and references therein).
% **WHY WAS THIS HERE?  (MALCOLM)** coherently oscillating spin-dependent effects due to the
%\note{BL: We should cite: The here described experiment obtained a new limit on axion-like particles using ultracold neutrons \cite{Afach2015Exotic}.}
%\note{YS: In the earlier (long) version of this paper, we actually had a reference to this paper and several other papers on related applications of ultracold neutrons to look for new physics. In the process of shortening the paper, the paragraph containing these references was cut out, but perhaps we could restore the references somewhere in the paper.}
Recently, however, it has been proposed to search for the interactions of the coherently oscillating axion DM field with gluons and fermions, which can induce oscillating electric dipole moments (EDMs) of nucleons \cite{Graham2011} and atoms \cite{Stadnik2014A,Roberts2014A,Roberts2014B}, and anomalous spin-precession effects \cite{Flambaum2013Patras,Stadnik2014A,Graham2013}.
The frequency of these oscillating effects is dictated by the axion mass, and more importantly, these effects scale linearly in a small interaction constant \cite{Graham2011,Stadnik2014A,Roberts2014A,Roberts2014B,Flambaum2013Patras,Graham2013}, whereas in previous axion searches, the sought effects scaled quadratically or quartically in the interaction constant \cite{Axion-Review2015}.
%\note{PMM: needs citation, even if repeated with [27-31]}

In the present work, we focus on the axion-gluon and axion-nucleon couplings:
\begin{align}
\label{Axion_couplings}
\mathcal{L}_{\textrm{int}} = \frac{C_G}{f_a} \frac{g^2}{32\pi^2} a G^{b}_{\mu \nu} \tilde{G}^{b \mu \nu}  - \frac{C_N}{2f_a} \partial_\mu a ~ \bar{N} \gamma^\mu \gamma^5 N \, ,
\end{align}
where $G$ and $\tilde{G}$ are the gluonic field tensor and its dual, $b=1,2,...,8$ is the  color index, $g^2 / 4 \pi$ is the color coupling constant, {\color{black}$N$ and $\bar{N} = N^\dagger \gamma^0$ are the nucleon field and its Dirac adjoint,} $f_a$ is the axion decay constant, and $C_G$ and {\color{black}$C_N$} are model-dependent dimensionless parameters.
Astrophysical constraints on the axion-gluon coupling in (\ref{Axion_couplings}) come from Big Bang nucleosynthesis \cite{Blum2014,StadnikThesis,Stadnik2015D}:~$m_a^{1/4} f_a / C_G \gtrsim 10^{10}~\textrm{GeV}^{5/4}$ for $m_a \ll 10^{-16}~\textrm{eV}$ and $m_a f_a / C_G \gtrsim 10^{-9}~\textrm{GeV}^{2}$ for $m_a \gg 10^{-16}~\textrm{eV}$, assuming that axions saturate the present-day DM energy density, %$\bar{\rho}_\textrm{DM} = 1.3 \times 10^{-6}~\textrm{GeV/cm}^3$ \cite{Planck2015},
and from supernova energy-loss bounds \cite{Graham2013,Raffelt1990Review}:~$f_a / C_G \gtrsim 10^6 ~\textrm{GeV}$ for $m_a \lesssim 3 \times 10^{7}~\textrm{eV}$.
{\color{black}Astrophysical constraints on the axion-nucleon coupling in (\ref{Axion_couplings}) come from supernova energy-loss bounds \cite{Raffelt1990Review,Raffelt2008LNP}:~$f_a / C_N \gtrsim 10^9 ~\textrm{GeV}$ for $m_a \lesssim 3 \times 10^{7}~\textrm{eV}$, while existing laboratory constraints come from magnetometry searches for new spin-dependent forces mediated by axion exchange \cite{Romalis2009_NF}:~$f_a / C_N \gtrsim 1 \times 10^4 ~\textrm{GeV}$ for $m_a \lesssim 10^{-7}~\textrm{eV}$. }

The axion-gluon coupling in (\ref{Axion_couplings}) induces the following oscillating EDM of the neutron via a chirally-enhanced 1-loop process~%\footnote{Interaction (\ref{Axion_couplings}) also non-perturbatively induces a mass $m_a \approx 6C_G\,\mu\text{eV} \cdot (10^{12}\text{ GeV}/f_a)$.
%Axions with masses much smaller than this are theoretically fine-tuned.}
\cite{tuningfootnote,Witten1979,*Witten1979B,Pospelov1999}:
%\note{KK: merge to [42-44]}
\begin{equation}
\label{eq:nEDM_axion}
d_\mathrm{n}(t) \approx +2.4 \times 10^{-16} ~ \frac{C_G a_0}{f_a} \cos(m_a t) ~ e \cdot \textrm{cm} \, .
\end{equation}
The axion-gluon coupling in (\ref{Axion_couplings}) also induces oscillating EDMs of atoms via the 1-loop-level oscillating nucleon EDMs and tree-level oscillating P,~T-violating intranuclear forces (which give the dominant contribution) \cite{Stadnik2014A,Flambaum1984EDM,*Flambaum1984EDMB}.
In the case of $^{199}$Hg, the oscillating atomic EDM is \cite{Stadnik2014A,StadnikThesis,Flambaum1985EDM,Flambaum1985EDMB,Flambaum2002EDM,Dmitriev2003A,Dmitriev2003B,Dmitriev2005,Engel2005,Engel2010}:
\begin{equation}
\label{199Hg-EDM_axion}
d_{\textrm{Hg}}(t) \approx +1.3 \times 10^{-19} ~ \frac{C_G a_0}{f_a} \cos(m_a t) ~ e \cdot \textrm{cm} \, ,
\end{equation}
which is suppressed compared to the value for a free neutron (\ref{eq:nEDM_axion}), as a consequence of the Schiff screening theorem for neutral atoms \cite{Schiff1963}.
%\note{KK: a question to our theory friends: Eqns. (2) and (3) suggest that $d_n$ and $d_\textrm{Hg}$ would have the same sign. Is this intended or are we rather talking about the modulus?}
%\note{YS: Yes, the same signs for $d_n$ and $d_\textrm{Hg}$ are intended here.}
%\note{MR: Should we explicitly point that out then?}
%\note{YS: I think this would be a good idea. We could explicitly add "+" signs in both Eqs. (2) and (3) to remove any ambiguity.}
%\note{CG: Might be good to comment that the Hg contribution is quite safely assumed to be negligible compared to the neutron for the purposes of this paper.}
The amplitude of the axion DM field, $a_0$, is fixed by the relation $\rho_a \approx m_a^2 a_0^2 /2$.
In the present work, we assume that axions saturate the local cold DM energy density $\rho_{\rm DM}^{\rm local} \approx 0.4~\textrm{GeV/cm}^3$ \cite{Catena2010}.

%Crucially, the constant amplitude of the axion DM field, $a_0$, is fixed by the known DM density at the location of the Earth, $\rho_{\rm DM}=\approx 0.4~\textrm{GeV/cm}^3$ \cite{Catena2010} (when reporting constraints on the couplings we assume axions make up all the local DM).

% Our experiment is also sensitive to the time-dependent energy shifts induced by the derivative coupling of an oscillating galactic axion DM field, $a = a_0 \cos(m_a t - \v{p}_a \cdot \v{r})$, with spin-polarised nucleons in (\ref{Axion_couplings}):
The derivative coupling of an oscillating galactic axion DM field, $a = a_0 \cos(m_a t - \v{p}_a \cdot \v{r})$, with spin-polarized nucleons in (\ref{Axion_couplings}) induces time-dependent energy shifts according to:
\begin{equation}
\label{potential_axion-wind}
H_{\textrm{int}} (t) = \frac{C_N a_0}{2 f_a} \sin(m_a t) ~ \v{\sigma}_N \cdot \v{p}_a \, .
\end{equation}
The term $\v{\sigma}_N \cdot \v{p}_a$ is conveniently expressed by transforming to a non-rotating celestial coordinate system (see, e.g., \cite{Kostelecky1999}):
\begin{align}
\label{sigma-p_a_2}
\v{\sigma}_N \cdot \v{p}_a  &= \hat{m}_F f(\sigma_N) m_a |\v{v}_a|  \notag \\
& \times \left[\cos(\chi) \sin(\delta) + \sin(\chi) \cos(\delta) \cos(\Omega_{\textrm{sid}} t - \eta) \right] \, ,
\end{align}
where $\chi$ is the angle between Earth's axis of rotation and the spin quantization axis ($\chi = 42.5 ^\circ$ at the location of the PSI), $\delta \approx -48 ^\circ$ and $\eta \approx 138 ^\circ$ are the declination and right ascension of the galactic axion DM flux relative to the Solar System \cite{NASA2014web}, $\Omega_{\textrm{sid}} \approx 7.29 \times 10^{-5}~\textrm{s}^{-1}$ is the daily sidereal angular frequency, $\hat{m}_F = m_F / F$ is the normalized projection of the total angular momentum onto the quantization axis, and $f(\sigma_N) = +1$ for the free neutron, while $f(\sigma_N) = -1/3$ for the $^{199}$Hg atom in the Schmidt (single-particle) model.

Here, we report on a search for an axion-induced oscillating EDM of the neutron (nEDM) based on an analysis of the ratio of the spin-precession frequencies of stored ultracold neutrons and $^{199}$Hg atoms, which is a system that had previously also been used as a sensitive probe of new non-EDM physics \cite{Altarev2009,Altarev2010,Afach2015_NF}.
We divided our analysis into two parts.
We first analyzed the Sussex--RAL--ILL nEDM experiment data \cite{Baker2014}, covering oscillation periods longer than days (\emph{long time--base}).
Then we extended the analysis to the data of the PSI nEDM experiment \cite{Baker2011}, which allowed us to probe oscillation periods down to minutes (\emph{short time--base}).
Our analysis places the first laboratory constraints on the axion-gluon coupling.
We also report on a search for an axion-wind spin-precession effect, using the data of the PSI nEDM experiment. Our analysis places the first laboratory constraints on the axion-nucleon coupling from the consideration of an effect that is linear in the interaction constant.

\section{Long time-base analysis}
The Sussex--RAL--ILL room temperature nEDM experiment ran from 1998 to 2002 at the PF2 beamline at the Institut Laue-Langevin (ILL) in Grenoble, France.
This experiment set the current world-best limit on the permanent time-independent neutron EDM, published in 2006 \cite{Baker2006, *Baker2006B}. The data were subsequently reanalyzed to give a revised limit in 2015 \cite{Pendlebury2015}. The technical details of the apparatus are described in full in \cite{Baker2014}, but we summarize the main experimental details here for the reader.

The experiment was based on Ramsey interferometry \cite{Ramsey1950} of ultracold neutrons \cite{UCN-Review1979,UCN-Review1994}.
%\note{CG: Might be good to define UCN or at least give a reference.}
The neutrons were stored in parallel or antiparallel electric and magnetic fields, where their Larmor precession frequency is given by
\begin{equation}
  \label{eq:Larmor}
  h\nu_\mathrm{n} = 2 \left| \mu_\mathrm{n} B \pm d_\mathrm{n} E \right| \, ,
\end{equation}
%\note{EW: I might be misremembering conventions, but shouldn’t the $n$ in $d_n$ and $mu_n$ be roman instead of italic? Same for the subscript a for axion? Or is the subscript a referring to the classical field a (in which case it should be italic)?}
%\note{YS: I don't think there's a standard convention on this. I've seen both italicised and Roman variants for elementary-particle subscripts in the literature. As long as we're consistent, then it shouldn't matter which convention we use.}
with the sign depending on the field configuration.
$E$ and $B$ are the magnitudes of the electric and magnetic fields, respectively.
By measuring the frequency difference between the two field configurations, a value for the neutron EDM, $d_\mathrm{n}$, was inferred.

%\note{PH: An irritant from the language perspective is the use of the "historic present" -- the measurement "is" rather than "was" conducted, and so on.}
%\note{PSW: I support PH remark, and would prefer a description of the experimental and analysis procedures in simple past (was instead of is).}
The measurement was conducted in a series of \emph{cycles}, each approximately 5 minutes long. A cycle began with a filling of neutrons polarized along the fields into the precession chamber from the ultracold neutron source~\cite{Steyerl1986}.
% \note{CG: Unclear to the reader what "the source" refers to.}
% Once they were within the chamber, an oscillating $\pi/2$ pulse \note{PSW: In general these pulses are far off from pi/2. May be call it a spin-tip pulse.} lasting 2 seconds was applied to begin the precession. Simultaneously, a population of polarized $^{199}$Hg atoms was released into the chamber, and a rotating $\pi/2$ pulse was also applied to set these atoms into free precession. The particles were stored and allowed to precess freely for 130 seconds. After this, the neutrons were subjected to a second $\pi/2$ analysing pulse.
Once they are in the chamber, enclosed from top and bottom with electrodes, a \unit[29]{Hz} NMR pulse lasting 2 seconds was applied to rotate the neutron spins into the transverse plane of the electromagnetic fields where they began to precess. Prior to the pulse, a population of polarized $^{199}$Hg atoms was released into the chamber, another 2-second \unit[29]{Hz} pulse, in-phase with the first one, was applied.
% d an \unit[8]{Hz} NMR pulse was applied to set these atomsorientation free precession. The particles were stored and allowed to precess freely for 130 seconds. After this, the neutron spins were rotated back into the longitudinal plane by a second \unit[29]{Hz} pulse.
%(ILL) or 180 seconds (PSI) \note{MR: not mention PSI here?}.
% \note{GP, TL, DR: "simultaneously, a population of polarized 199Hg atoms is released..." -> "Prior to the pi/2 pulse, a population...", "precess freely for 130 s" -> it is in fact 180s for the PSI data. }
%\note{FP: you call the second pi/2-pulse "analysing pulse". I guess this is somehow confusing (and maybe formally not correct), especially since one line afterwards you mention the "analyzing foil"}.
The neutrons were then emptied into a detector through a spin-analyzing foil. Over 1--2 days, many of these cycles were performed. The electric field's polarization was reversed every hour. We term one continuous block of data taking in the same magnetic-field configuration, but including both directions of electric field, a \emph{run}. One run gives a $d_\mathrm{n}$ estimate.
%\note{MR: This would be the place to mention the crossing lines, if we want it. Maybe a short comment about the one-run ILL $d_n$ estimate would be in place. Does it still include systematic effects? Or does it not? Later, we say that we assume zero offset in fit, it may need a little motivation.}

In order to suppress cycle--to--cycle changes in the magnetic field, the analysis was performed on the ratio of the neutron and mercury precession frequencies $R$, which, using (\ref{eq:Larmor}), is \cite{Baker2014}:
\begin{align}
  \label{eq:R}
  % R &= \frac{\nu_n}{\nu_{Hg}} = \frac{\mu_n}{\mu_{Hg}} \pm \left( d_n - \frac{\mu_n}{\mu_{Hg}} \, d_{Hg} \right) \frac{E}{\mu_{Hg} B} + \mathcal{O}(d^2) \ ,
  R &\equiv \frac{\nu_\mathrm{n}}{\nu_\textrm{Hg}} = \frac{\mu_\mathrm{n}}{\mu_\textrm{Hg}} \pm \left( d_\mathrm{n} - \frac{\mu_\mathrm{n}}{\mu_\textrm{Hg}} \, d_\textrm{Hg} \right) \frac{2 E}{ h  \nu_\textrm{Hg}} + \Delta \, ,
\end{align}
where the signs correspond to parallel and antiparallel field configurations. $\Delta$ encapsulates all higher-order terms and systematic effects, which are corrected for when a run is analyzed~\cite{Pendlebury2015}.
% \note{FP: what is the $d^2$ in the higher order term of the equation - is it a dipole moment? But then it has no index (n or Hg), if it's something else it is not explained I guess.}
% \note{MR: We need $\mathcal{O}(d^2)$, as there are terms $\propto d_n^2$, $\propto d_{Hg}^2$, and $\propto d_n d_{Hg}$. Will need to clarify that in words, I can't come up with a sensible notation.}
% \note{PSW: You do not only have corrections of higher order of electric-dipole signals but also due to the different adiabatic regimes. Try to find a better way of generalizing higher order terms.}
This analysis is sensitive to oscillations in the quantity $d_\mathrm{n} - \left( \mu_\mathrm{n} / \mu_\textrm{Hg} \right) \, d_\textrm{Hg}$, with $\mu_\mathrm{n} / \mu_\textrm{Hg} = -3.8424574(30)$~\cite{Afach2014magmoment}.

%, which we conveniently give in $\unit[10^{-26}]{e \cdot cm}$ units. %\note{MR: Need to give the magnetic field strength ($B$ in the above equation) used for the $R$ to $d_n$ conversion. For the PSI analysis is taken to be $\unit[1.03536]{\upmu T}$.}
% \note{EW: I think equation 5 is incomplete. In principle it should also include all systematic effects influencing R, such as
% - rotation of the earth
% - $<Bt^2>$
% - Linear and cubic GPE in fHg/fn
% - HgM light shift
% - Shift due to gravity (G1,0 and G3,0 mainly, others as well)
% - …?
% Seeing that most of them aren’t really relevant to the story, maybe you can condense them into a delta or something? But I think you should at least mention the shift due to gravity, as this one reappears a bit later in the paper. (And you mention the crossing lines procedure as well.)}
% \note{YS: Maybe could introduce $\Delta$ as the last term in (5), defined as ``all higher-order terms and systematic effects (such as the slight downwards displacement of the ultracold neutrons with respect to the hot cohabitating mercury atoms'' [Give a reference here?])?}

In our analysis, we were looking for an oscillating EDM. We performed this search in frequency space by evaluating \emph{periodograms} -- estimators of the power spectrum. An oscillation in the time domain would show up as an excess in the power (or, equivalently, amplitude) relative to the expected distribution due to experimental noise.

\begin{figure}
  \centering
  \includegraphics[width=\columnwidth]{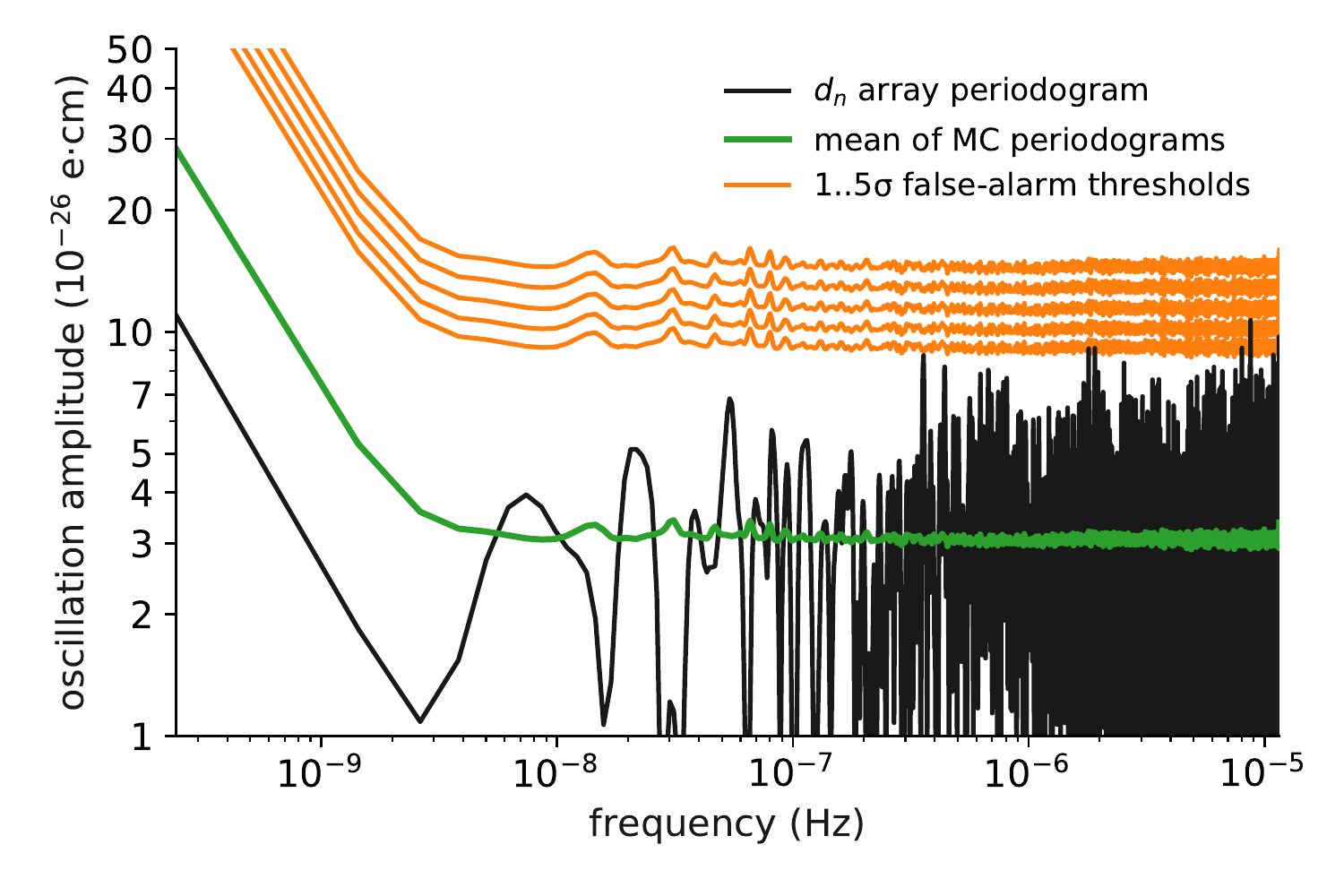}
  \caption{The periodogram of the array of neutron EDM ($d_\mathrm{n}$) estimates from the ILL measurement (black line).
  We are sensitive to oscillations in the quantity $d_\mathrm{n} - \left( \mu_\mathrm{n} / \mu_\textrm{Hg} \right) \, d_\textrm{Hg}$, where $d_\textrm{Hg}$ is the EDM of the $^{199}$Hg atom. The mean of Monte Carlo (MC)-generated periodograms, assuming no signal is present, is depicted in green. MC is used to deliver false--alarm thresholds (global $p$-values), marked in orange for $1,2,…,5\,\sigma$ levels (from bottom to top). The highest peak has the global $p$\-/value 0.53, consistent with a non-detection.}
  \label{fig:ILL_detection}
\end{figure}

In the case of the long time--base analysis, we considered the time series of $d_\mathrm{n}$ measurements from individual runs (after having corrected for the ``False EDM'' effect \cite{Pendlebury2004} using the crossing lines procedure \cite{Pendlebury2015}). %\note{NA: mention crossing lines here, because it doesn't make sense until you explain that you use the R rather than $\nu_{n}$},
The measurements are neither evenly spaced nor have equal uncertainties.
%To calculate the periodogram of the data series, we use the Least Squares Spectral Analysis (LSSA) \cite{Scargle1982,Cumming2004}, where the amplitude at frequency $f$ is estimated by the amplitude of the best fit oscillation of that frequency. We evaluate the periodogram at a set of 1334 \note{@NA todo double check} trial frequencies, evenly spaced between 0 (exclusive) and $\unit[10]{\upmu Hz}$ \note{@Nick verify}. The spacing, $\unit[7.49]{nHz}$, is chosen to be equal to the intrinsic spectral resolution of the dataset (its inverse time span).
%An axion DM signal (with expected coherence set by $\Delta \omega/ \omega \sim 10^{-6}$) is narrower than the spectral resolution in the whole range of tested frequencies. The width of a signal peak is thus set by the spectral resolution, which we choose to be the spacing of the trial frequencies.
%By doing so, we ensure that we do not miss a DM signal in the entire considered frequency range.
To calculate the periodogram of the data series, we used the Least Squares Spectral Analysis (LSSA) method \cite{Scargle1982,Cumming2004}, where the amplitude at frequency $f$ was estimated by the amplitude of the best fit oscillation of that frequency. We evaluated the periodogram at a set of 1334 trial frequencies, evenly spaced between \unit[100]{pHz} (arbitrarily chosen, a period of about 300 years, much longer than the four--year span of the data set) and $\unit[10]{\upmu Hz}$ (a period of about a day, the time it typically took to get one $d_\mathrm{n}$ estimate).
An axion DM signal, with expected coherence set by $\Delta f \sim 10^{-6} f~$\footnotemark[1],
% \note{PH: not clear where it's from.} \note{YS: Added a reference to the existing footnote to clarify this point.}
is narrower than the spectral resolution ($\unit[7.49]{nHz}$, the inverse span of the data set) in the whole range of frequencies we were sensitive to.
% \note{FP: This is a bit confusing to me since you compare relative with absolute resolution. Can you maybe in addition translate the \unit[7.5]{nHz} to a relative resolution.}
% \note{MR: The key is \emph{the whole range of frequencies}; the relative relation holds in the whole range. Formulate better.}
% \note{YS: Maybe we could rewrite $\Delta f/ f \sim 10^{-6}$ as $\Delta f \sim 10^{-6} f$ above?}
% The width of a signal peak is thus set by the spectral resolution, which we chose to be the spacing of the trial frequencies. We evaluated the periodogram at a set of 1334 trial frequencies, evenly spaced between \unit[100]{pHz} (arbitrary chosen, a period of about 300 years, much longer than the span of the data set) and \unit[10]{\upmu Hz} (a period of about a day, the minimal separation between  $d_n$ estimation in the ILL experiment).
%evenly spaced between $\unit[100]{pHz}$ (arbitrarily chosen) and $\unit[10]{\upmu Hz}$.
% \note{MMM: Was 10 uHz also arbitrarily chosen or is there a reason for it?}
% \note{JK: Evenly in linear or log-scale? Why this choice?}
% \note{JZ: evenly spaced between 100 pHz (arbitrary chosen, period of about 300 years which much longer than experiment duration) and 10 uHz (period of about 1 day - minimal time of estimation of $d_n$ in the ILL experiment)}
% \note{@Nick: Needs clarification which set of frequencies is used for the analysis. The text and plots need to be consistent.}
In the LSSA fit, we assumed the free offset to be zero on the grounds that the experiment had already delivered a zero-compatible result for the permanent time-independent neutron EDM \cite{Baker2006, *Baker2006B,Pendlebury2015}. The periodogram of the long time--base dataset is shown as a black line in Fig.\,\ref{fig:ILL_detection}.
% \note{DM: The rise in the ILL analysis occurs near 7.49 nHz as inverse span of dataset. This is fine enough explanation for me. Mark as point on Fig. 2?}
% \note{GP, TL, DR: Then we expect some kind of comment about it. Maybe some more pedagogy is needed for the next paragraph. Is this necessary to give the technicalities about the extrapolation of the CDF, maybe just give the reference? } To estimate the expected distribution of the periodogram, we used Monte Carlo (MC) simulations. We generated 5000 datasets, assuming no signal is present, and evaluated the periodogram of each.
%The area covered by the generated periodograms is shown in Fig.\,\ref{fig:ILL_detection} in form of green bands.
To obtain the expected distribution of the periodogram, we performed Monte Carlo (MC) simulations. At each frequency, we estimated the cumulative distribution function (CDF) of the LSSA power. Extreme events in the tails of the distribution are expensive to access directly with MC. For this reason, to the discrete CDF estimates we fitted, at each $i^\textrm{th}$ frequency, the functional form of the LSSA-power CDF~\cite{Scargle1982}:
\begin{equation}
	\label{eq:powerdistribution}
	F_i(\mathcal{P}) = 1 - A_i\,\exp(-B_i\, \mathcal{P}) \, ,
\end{equation}
% \note{JZ: I do not like equations without "=" sign. Maybe it will be better to write $F(P) = 1-Aexp(BP)$}
where $\mathcal{P}$ is the power, while $A_i$ and $B_i$
% \note{PMM: B is already used for magnetic field before, please use something else here.}
% \note{YS: We could in principle use the mathcal variants of `A' and `B' here, but I don't think a reader would confuse the 'B' here for the magnetic field.}
are fit parameters. The local $p$-values are given by
\begin{equation}
	\label{eq:localpvalue}
    p_{\mathrm{local}, i} = 1 - F_i(\mathcal{P}_i) \, ,
\end{equation}
where $\mathcal{P}_i$ is the LSSA power of the measured $d_\mathrm{n}$ time series at the $i^\textrm{th}$ frequency.
% \note{CG: awk., a bit of a comma run-on} At each frequency, comparing the best-fit amplitude in the real dataset to its extrapolated CDF gives the local $p$\-/values. We find the minimum local $p$\-/value to be $7.3 \times 10^{-4}$ at the frequency of $\unit[8.02]{\upmu Hz}$, which corresponds to 1.44 inverse days.

If the local $p$\-/values at different trial frequencies were uncorrelated, the global $p$\-/value would be given by \cite{Algeri2016}:
\begin{equation}
	\label{eq:pvalues}
    p_\mathrm{global} = 1 - (1 - p_\mathrm{local})^N \, ,
\end{equation}
where $N$ is the number of trial frequencies.
% \note{YK: At Eq. (7), the discussion on how to extract the minimal local p-value wouldn't be clearer with a plot like slide 29 of your Bern presentation? I'm asking because this looks like as a part of the core of the analysis and afterwards you discuss in more details the penetration of the thresholds. Graphical explanation on that point would have helped me to get the estimation of the false alarm thresholds (also, I would understand if this kind of plot is too detailed\ldots).}
However, we did not need to make this assumption.
% \note{YK: In the paper N(best) = 1026 while N(trials) = 1334. 30\% difference in N translate in a "huge" difference in the CDF if I plot it, isn't it? -> Does that mean we have "huge" correlation of R fluctuations in between different frequencies? -> If so, why? I'm not sure to understand what in our case, could make the local p-values correlated in the frequency domain. Maybe give examples to illustrate why you "do not feel safe"...}
Instead, we made use of the set of MC datasets.
In each, we found the minimal local $p$\-/value and estimated its CDF, assuming it has the form (\ref{eq:pvalues}), but left $N$ as a free parameter.
We found the best fit value $N_\mathrm{effective} = 1026$. %\note{YS: Is the last digit significant here?} \note{NA: no it is not significant, however I wanted it to be clear we do not require a whole number. It would be okay to omit .1 .}
For each frequency, we marked the power necessary to reach the global $p$\-/values corresponding to $1,2,…,5\,\sigma$
% \note{FP: maybe change to "$1,2,…,5\,\sigma$"}
% \note{EW:  Make sure that 1..5sigma is written the same way in the text and the caption and legend of Fig 1 and 3.}
levels as orange lines in Fig.\,\ref{fig:ILL_detection}. The minimal local $p$\-/value of the dataset translates to the global $p$\-/value of 0.53, consistent with a non-detection.
%\note{GP, TL, DR: This value seems very large, no? Is it not 0.053 instead?}

\begin{figure}
  \centering
  \includegraphics[width=\columnwidth]{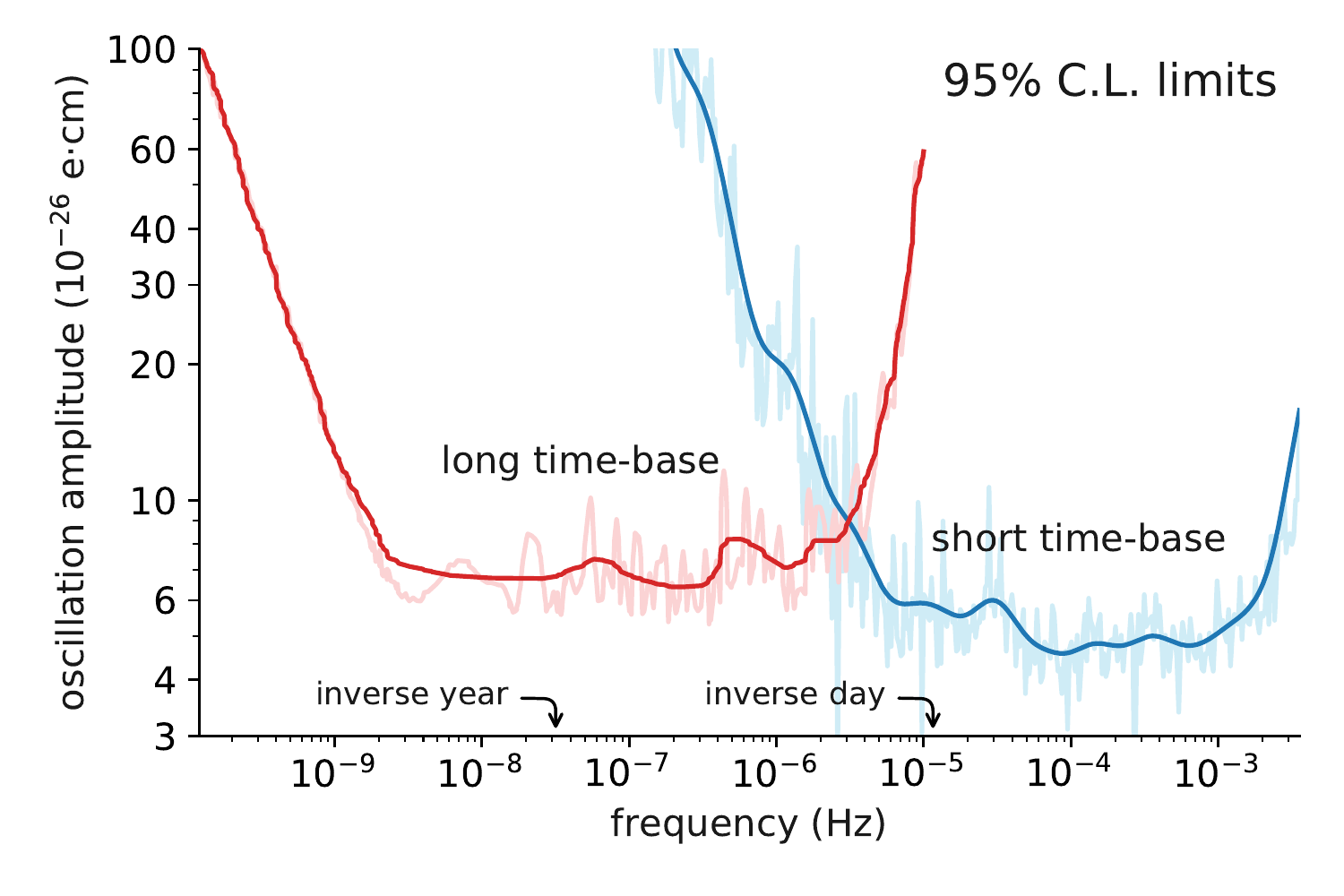}
  \caption{The 95\% C.L. limits on the amplitude of oscillation in the quantity $d_\mathrm{n} - \left( \mu_\mathrm{n} / \mu_\textrm{Hg} \right) \, d_\textrm{Hg}$, as a function of frequency thereof. The limits from the long (ILL data) and short (PSI data) time--base analyses are depicted by the red and blue curves, respectively, with the area above these curves being excluded.
  The raw limits delivered by the analysis, with substantial noise, are depicted by the light lines, while the smoothed versions are given in bold.
  }

  \label{fig:ecm_limits}
\end{figure}

In order to obtain limits on the oscillation amplitude parameter, we again used MC simulations. We discretized the space of possible signals, spanned by their frequency and amplitude. We chose a sparser set of 200 frequencies, as we did not expect highly coherent effects in the sensitivity of detection. For each discrete point, we generated a set of 200 MC
%\note{PMM: perhaps use the words 'toy' or 'sample' instead of fake}
datasets containing the respective, perfectly coherent signal and assumed that the oscillation is averaged over the duration of the run. In general, the sensitivity is phase-dependent, especially for periods comparable with the length of the dataset. For simplicity, we did not investigate the phase-dependence and in the simulation took it to be random and uniformly distributed.
%\note{NA: averaged over run (1 day ish) for ILL, one free precession time for PSI}
% \note{DM: do we need to note that this assumes that the coherence time is at least the length of the run? This is satisfied by the $\Delta f/f\sim 10^{-6}$ width.}
For each fake dataset, we evaluated the LSSA amplitude only at the frequency of the signal and compared its distribution (extrapolating with the functional form of Eq.\,(\ref{eq:powerdistribution})) with the best-fit amplitude in the data and defined the $p$\-/value to be left--sided. We found the 95\% confidence-level exclusion limit as the 0.05 isocontour of the $\mathrm{CL}_s$ statistic \cite{PDG2016}.
The limit is shown as the red curve in Fig.\,\ref{fig:ecm_limits}.
We are most sensitive to periods shorter than the timespan of the dataset ($\sim 4$ years), %\note{NA: we do have good sensitivity down to about 3*10-9Hz or 10 inverse years, but how to phrase this?} (\note{@Nick give a number}), %NA: sensitivity in e cm terms is roughly flat from longer than total measurement time up to individual data point time
% \note{CG: implies to me a 10 year dataset (1998-2002?)}
but rapidly lose sensitivity for periods shorter than the temporal spacing between data points ($\sim 2$ days), since the expected signal would essentially average to zero over these short time scales.

% \note{MR: Try to make the transition ILL -> PSI only once, so that the reader is not confused. Below first describe the changes made to the ILL apparatus and why, as motivation, they enable going to the short time--base analysis.}

\section{Short time-base analysis}
In 2009, the Sussex--RAL--ILL apparatus was moved to the new ultracold neutron source at the Paul Scherrer Institute (PSI), Villigen, Switzerland~\cite{Anghel2009,Lauss2011,Lauss2012,Lauss2014}, where a number of improvements were made~\cite{Baker2011,Afach2015USSA,Ban2016NANOSC}. In 2015, the apparatus was fully commissioned and began to take high-sensitivity EDM data. The whole data set, taken from August 2015 until the end of 2016, with a higher accumulated sensitivity than the ILL one, was considered in this analysis. For the PSI experiment's data, we performed a lower--level oscillation search on the array of $R$ measurements. Since an $R$ estimate was obtained every cycle ($\unit[\approx 300]{s}$), rather than every 1--2 days as for a $d_\mathrm{n}$ estimate, it has an increased sensitivity to higher frequencies. Additionally, the analysis could benefit from the addition of 16 atomic cesium vapor magnetometers \cite{Knowles2009, Koch2015},
% \note{KK: Add 2 references here: P. Knowles et al., NIMA 611, 306 (2009); S. Afach et al., EPJD 69, 225 (2015).}
located directly above and below the precession chamber (inside the electrodes). This made it possible to account for the dominant time--dependent systematic effect on a cycle, rather than run, basis.
% \note{GP: State already here, that there is higher sensitivity. This is the motivation to have two different analyses. In addition, we have the Cs magnetometers, which allows... (In ILL we saw, that there are no significant drifts of the gradient. Even if there are, these are accounted for by increasing the error-bar. The HV reversal is to be able to correct for gradient fluctuation.)}

The dominant time-dependent systematic effect, encapsulated in $\Delta$ of Eq.\,\eqref{eq:R}, would have given rise to non-statistical temporal fluctuations if not accounted for. Namely, $R$ is sensitive to drifts in the vertical gradients of the magnetic field. While the thermal mercury atoms filled the chamber homogeneously, the center of mass of the ultracold neutron population was lower by several millimeters~\cite{Afach2014magmoment, Afach2015, Pendlebury2015}.
%\note{KK: Add 2 more references here: S. Afach et al., PRL 115, 162502 (2015); S. Afach et al., PRD 92, 052008 (2015).}
% \note{MR: This can be added, if needed. The additional contribution to $R$ is: \note{MR: the explicit equation may be left out, if we need space.}
% \begin{equation}
%   \label{eq:deltah}
%   \Delta_{\Delta h} = \mp \Delta h \left[ \frac{\nu_n}{\nu_\textrm{Hg}} \,
%     \frac{1}{B} \left( G_{1,0} - \frac{3 r^2}{4} G_{3,0} \right) \right] \,
% \end{equation}
% where $G_{1,0}$ and $G_{3,0}$ are cubic and linear vertical gradients of the magnetic field \note{MR: cite the paper about the formalism (Guillaume Pignol}.}
To evaluate the correction, the drifts of the gradients were estimated on a cycle basis by fitting a second--order parametrization of the magnetic field to the measurements of the cesium magnetometers~\cite{WurstenThesis}. The center-of-mass shift was determined to be \unit[4]{mm} using the method described in \cite{Afach2014magmoment}.

The measurement procedure involved working deliberately with gradients affecting $R$ (see the crossing--point method in \cite{Pendlebury2015}). Those intended gradients (up to \unit[60]{pT/cm} in \unit[10]{pT/cm} steps) were much larger than cycle--to--cycle fluctuations (\unit[< 2]{pT/cm} per day). With the high-order shifts in $R$ having been significant, these large shifts could not be corrected using the cesium magnetometers. Additionally, while the cesium magnetometers were precise, their accuracy is limited by the calibration procedure. We defined as a \emph{sequence} a set of data, typically 2--3 days in duration, without a deliberate change in the magnetic-field gradient or a recalibration of the cesium magnetometers.
% On the other hand, the precise, but not accurate, cesium magnetometers can only measure the change of the gradient. We need to assume a new random offset in the measurement each time the magnetometers are recalibrated. \note{JK: Why do we recalibrate?} We define as a \emph{sequence} a set of data, typically 2--3 days in duration, without a renewal of the calibration.
% \note{EW: Regarding the sequences in the short time-base dataset. It is technically correct that each sequence is defined by a calibration of the CsMs, but even if we would not calibrate each time when we start a run, you would still have to split based on changes in the magnetic field configuration (R shifts due to G3,0, $<Bt^2>$ and GPE, which is not compensated for with the CsM correction).}
When performing the LSSA fit, we allowed the free offset to be different in each sequence:
\begin{equation}
	% A\sin(2 \pi f t) + B\cos(2 \pi f t) + C_1\,\Pi_1(t) + C_2\,\Pi_2(t) + \ldots \,
    A\sin(2 \pi f t) + B\cos(2 \pi f t) + \sum_i C_i\,\Pi_i(t) \, ,
\end{equation}
where $C_i$ is the free offset in the $i^\textrm{th}$ sequence and $\Pi_i(t)$ is a gate function equal to one in the $i^\textrm{th}$ sequence and zero elsewhere.
% \note{FP: I am a bit confused why have all the constants C in the equation. Should this not read like:  $A\sin(2 \pi f t) + B\cos(2 \pi f t) + C_i$  and then you fit each region with a different $C_i$. Now it is written as if all the offsets add up to a total offset of $C_1 + C_2 + C_3 + \ldots$ }
% \note{MR: Fair point. In fact each $C_i$ is actually $C_i \cdot g_i(t)$, where $g_i(t)$ is a gate function, one in $i$th sequence and zero elsewhere.}
This caused the short time--base analysis to lose sensitivity for periods longer than one sequence.
It should also be mentioned that, at the time of this analysis, the PSI data were still blinded, whereby an unknown, but constant, $d_\mathrm{n}$ was injected into them. It does not influence this analysis, as the free offsets are not considered further.

We split the $R$ time array into three sets:
% A control set is data taken without an applied electric field.
% There are two sets that are sensitive to an oscillating EDM, namely the sets with the applied electric and magnetic fields parallel and antiparallel.
a control set of data without an applied electric field, and two sets sensitive to an oscillating EDM, namely with parallel and antiparallel applied electric and magnetic fields.
% \note{CG: Might be good to say what the standard sequence is: 48 at HV, 8 at zero}
% \note{PSW: A control set of data without applied electric field, and two sets sensitive to an oscillating EDM, namely with parallel and antiparallel applied electric and magnetic fields.}
A coherent oscillating EDM signal would have an opposite phase in the latter two sets, and be absent in the control set.
We did not perform a common fit. Instead, the two sensitive data sets were treated separately in the LSSA fits, and later combined to a limit. Otherwise, the LSSA treatment was the same as in the long time--base analysis. We picked a set of $156\,198$ trial frequencies, spaced apart at intervals determined by the spectral resolution
%\note{PSW: Is that correct English? "spaced every spectral"}
%\note{YS: Maybe ``spaced apart at intervals determined by the spectral resolution''?}
(the inverse of 506 days = $\unit[23]{nHz}$), which here also defines the signal width.

\begin{figure}
  \centering
  \includegraphics[width=\columnwidth]{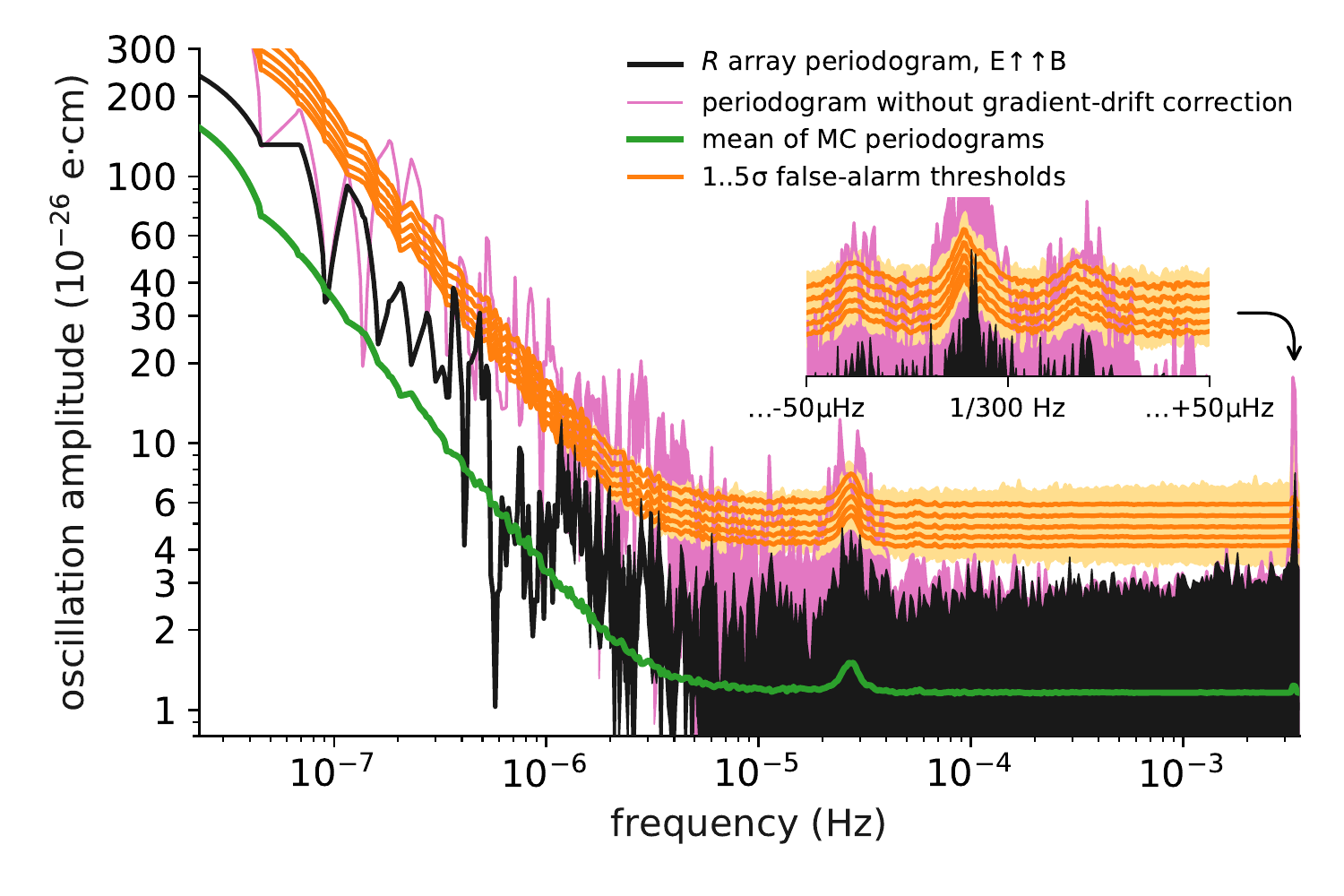}
  \caption{
  Periodogram of the $R$ time array of the PSI experiment data, sensitive to oscillations in the quantity $d_\mathrm{n} - \left( \mu_\mathrm{n} / \mu_\textrm{Hg} \right) \, d_\textrm{Hg}$, taken with the $\v{E}$ and $\v{B}$ fields parallel (black line).
  The mean of MC--generated periodograms, assuming no signal, is depicted in green. MC is used to calculate $1,2,…,5\,\sigma$ false--alarm thresholds, depicted in light orange.
  % \note{YS: It's a bit difficult to see the yellow on this graph. Or perhaps yellow means light orange here?}
  For clarity, we also plot the smoothed version in orange.
  %As they are noisy, a smoothed version is plotted on top in orange.
  There are two regions where a rise in the amplitude is expected, namely around $\unit[28]{\upmu Hz}$ (inverse of 10 hours) and $\unit[3.3]{mHz}$ (inverse of 300 seconds), due to the time structure of the data taking (see the main text for more details). The periodogram of non-gradient-drift-corrected data is shown in pink.
  % \note{MR: Maybe for the PRL it would be good not to show the narrow peak? Just to trim the frequency range a bit. It would save a lot of explanations, but I don't know if we can do that.}
  % \note{MR: comment on filtering of the limits and on the inset (remind the cycle repetition rate: 300s.}
%   \note{YK: how do you get periodogram values above 1/300s for the short-time base? (I would expect a really sharp end of the period. there)}
  % \note{JZ: in both the main text and figure caption you use 3.3 mHz. Maybe it will be less confusing to place in the x-axis of the inset 3.3 mHz instead of 1/300 Hz}
%   \note{JZ: one of three data sets is presented here. Which one? It is not written, I am afraid. I think it would be much more convincing if you will show Fig 3 at least for two data sets because you are not able to give exact reasons for some observed picks. There are some speculations only. If these picks exist in data sets collected with and without electric field it would helpful to understand these speculations. It is better to the reader to see it and not only to read about it.}
  }
  \label{fig:PSI_detection}
\end{figure}

The periodogram of the $R$ time array taken with the parallel-field configuration is shown in black in Fig.\,\ref{fig:PSI_detection}.
There are two regions of expected rise in the oscillation amplitude due to the time structure of the data collection.
The one around $\unit[28]{\upmu Hz}$ (the inverse of 10 hours) corresponds to the period of the electric-field reversal.
The very narrow one around $\unit[3.3]{mHz}$ (the inverse of $\unit[300]{s}$) corresponds to the cycle repetition rate.
There are five
%\note{NA does anybody know what style guide says about when numbers should be written in words i.e. seven vs 7? I personally think small numbers without units should be words but this comes down to preference} ***I THINK LESS THAN TEN IT IS IN WORDS - MALCOLM ***
trial frequencies for which the $3\sigma$ false--alarm threshold is exceeded,
%\note{PMM: use words like 'surpassed', or 'exceeded' instead of penetrated}
two of which, including the largest excess with a $6\sigma$ significance, occur in a $\unit[100]{\upmu Hz}$ region around the inverse of \unit[300]{s}, while the other three are in the low-frequency region (inverse days) already excluded by the long time--base analysis.
% \note{FP: Can you maybe also explicitly state the positions of these three lines in Hz or inverse time?}
 The periodograms for the other two datasets (not shown) are very similar.
In the other sensitive set, there are three excesses of the $3\sigma$ threshold (the highest is $5\sigma$), all constrained to the same two regions. In the control dataset, only the $1\sigma$ threshold is exceeded.
%\note{MR: We may consider not discussing the false--alarm threshold penetrations, and just state in the next paragraph that we do not find a signal which meets our criteria.}
The periodogram of the $R$ time array without the gradient-drift correction is shown in pink in Fig.\,\ref{fig:PSI_detection} to visualize the frequencies where the correction has an effect.

A non--statistical excess in a periodogram of $R$ may be caused not only by a coherent oscillating signal;
for example, fluctuations of a higher--order term in the magnetic field, not compensated by either the mercury or cesium magnetometers, may cause broad--band elevations in LSSA power.
% \note{MMM: The previous sentence is long and hard to follow.  I suggest rewording.}
% This was not a problem in the long time--base analysis, where the manner in which each $d_n$ estimate was obtained ensured correlation with the relative orientation of the electric and magnetic fields, rejecting all uncorrelated fluctuations.
% \note{MR: explain that the data is low level. Point to a possible source of non-statistical behaviour? E.g. In particular, a careful study of the Caesium magnetometer array is still ongoing. Or say that the analysis is ongoing. But not give the impression that we show a work-in-progress analysis.}
We defined strict requirements for an excess to be considered as one induced by axion DM as follows.
Firstly, a significant ($>3\sigma$) excess in amplitude had to be observed in both sensitive datasets at the same frequency, but not in the control set. Secondly, the signals must be in antiphase in the parallel and anti-parallel datasets. Lastly, we require high coherence (a narrow peak) equal to the spectral resolution of the dataset. None of the significant excesses passed our discovery criteria.
%\note{DM: did you also need to use the fact that some peaks were excluded by the long-time base analysis, or does the PSI data independently exclude all of the excesses based on these criteria? MR says PSI independently excludes peaks so make sure we say this.}
% \note{GP, TL, DR:
% In the paragraph where we apologize for these peaks, there is a sentence "This was not a problem in the long time-base analysis, where the manner in which each $d_n$ estimate was obtained ...." We believe it is not correct.  For high frequency noise (higher than the E field reversal frequency) the averaging over the cycles do not suppress the extra noise.  In this case we would see an increased of the power in the periodogram for each frequency. We suggest to remove this sentence.
% Beside giving excuses, what did we do to try to understand the origin of these significant peaks?
% To go in this direction we insist it is of utmost importance to perform the periodogram analysis of the west timeseries to see if these peaks are there too. }
% \note{MR: The analysis of the Western data is ready, waining for response...}

We delivered a limit on the oscillation amplitude similarly to the long time--base analysis, with the exception that we required the product of the two sensitive sets' $\mathrm{CL}_s$ statistics to be 0.05.
The limit is shown as the blue curve in Fig.\,\ref{fig:ecm_limits}.
With the short time--base analysis, we were most sensitive to periods shorter than the timespan of a sequence (2 -- 3 days), and lost sensitivity to periods shorter than the cycle repetition rate ($\approx 5$ minutes).
The PSI dataset has a higher accumulated sensitivity than the ILL dataset, so the limit baseline in the sensitive region is slightly better in the case of the PSI dataset.

Following Eq.\,(\ref{eq:nEDM_axion}), we can interpret the limit on the oscillating neutron EDM as limits on the axion--gluon coupling in Eq.\,(\ref{Axion_couplings}).
We present these limits in Fig.\,\ref{fig:axion_limits_v2}, assuming that axions saturate the local cold DM energy density $\rho_{\rm DM}^{\rm local} \approx 0.4~\textrm{GeV/cm}^3$ \cite{Catena2010}.
%\note{GP: DM experiments chose to use the value of 0.3 GeV, and it is pessimistic on purpose. Need to answer! }
%\note{YS: The current (and several previous) Particle Data Group reviews on DM have consistently quoted the value $0.4~\textrm{GeV}/\textrm{cm}^3$, presumably because this value represents the mean of the various independent determinations of the DM densities in our local galactic neighbourhood. I'm not saying that $0.3~\textrm{GeV}/\textrm{cm}^3$ is an incorrect value (after all, it lies within the range of possible values), but I think the pessimistic/conservative approach is not justified in the present case, because from independent observations of planetary orbits within the Solar System (which importantly do not contradict larger-scale determinations of the DM density) the density of DM on Earth and hence in laboratory experiments can be up to 6 orders of magnitude larger than the quoted value $0.4~\textrm{GeV}/\textrm{cm}^3$; see, e.g., https://arxiv.org/abs/astro-ph/0601422 }
Our peak sensitivity is
%\note{YS: Based on Fig.~4, I would have expected something closer to $f_a/C_G \approx 10^{21}$ GeV}
% \note{YS: The figure indicates a number close to $f_a/C_G \sim 10^{21}$ GeV, rather than $10^{-21}$ GeV$^{-1}$, which is for the reciprocal quantity $C_G/f_a$. It seems reasonable to give the sensitivity here to 1 digit of significance, i.e. $f_a/C_G \approx y \times 10^{zz}$ GeV. }
$f_a/C_G \approx \unit[1 \times 10^{21}]{GeV}$ for $m_a \lesssim \unit[10^{-23}]{eV}$, which probes super-Planckian axion decay constants ($f_a > M_{\textrm{Planck}} \approx 10^{19}~\textrm{GeV}$), that is, interactions that are intrinsically feebler than gravity.
%\note{JK: axion coupling does go $1/r^2$?}\note{DM: yes, but it also has a term proportional to the dot-product of the spins, and a Yukawa suppression for the axion mass. Here I read this as referring rather to the energy scale of suppression of terms in an effective field theory, regardless of the specific type of force they mediate.}
%\note{YS: We could in principle avoid having to say most of this, if we follow your (and my) suggestion in the figure caption to simplify the constraints graph to include only our nEDM results and the existing BBN constraints. If we do this, then we should still explain the $\sqrt{\rho}$ scaling of the constraints (either in the caption or in the text).}
%\note{YS: By the way, Doddy, where do you see the 4 orders of magnitude improvement in Fig.~4? (It's possible that we could be seeing different renderings of the graph in our viewers, since for me the dotted line corresponding to the Planck scale at $f_a \approx 10^{19}$ GeV appears to be shifted a bit upwards compared with where I would expect to see it. In this case, we should compare the nEDM and BBN limits using the raw number from the [@Nick] part above.}

% To search for the axion-wind effect, Eq.~(\ref{potential_axion-wind}), we partitioned the entire PSI dataset into two sets with opposite magnetic-field orientations (irrespective of the electric field) and then analyzed the ratio $R = \nu_\mathrm{n} / \nu_\textrm{Hg}$ similarly to our oscillating EDM analysis above.

\section{Axion-wind effect}
\begin{figure}
  \centering
  \includegraphics[width=\columnwidth]{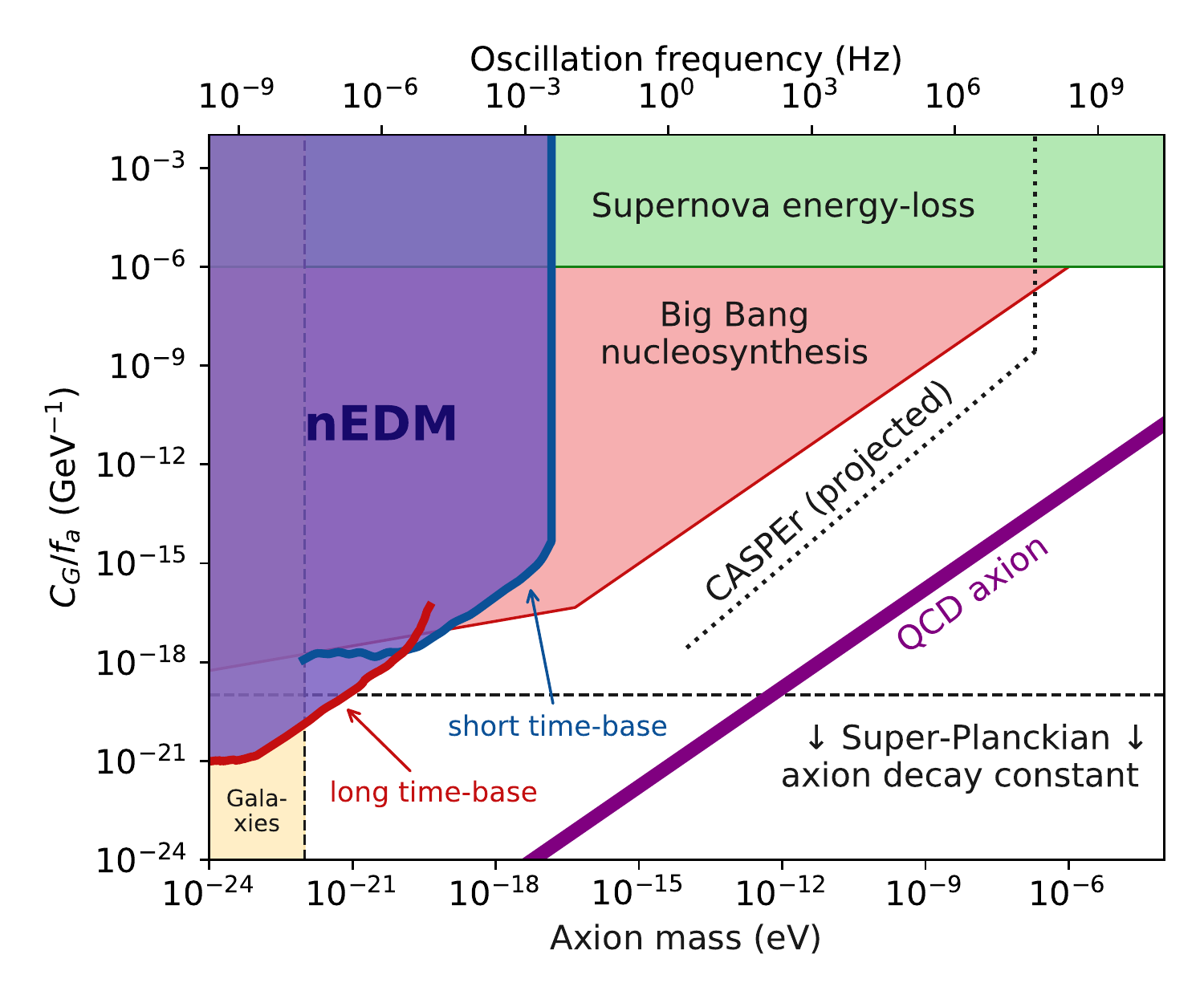}
  \includegraphics[width=\columnwidth]{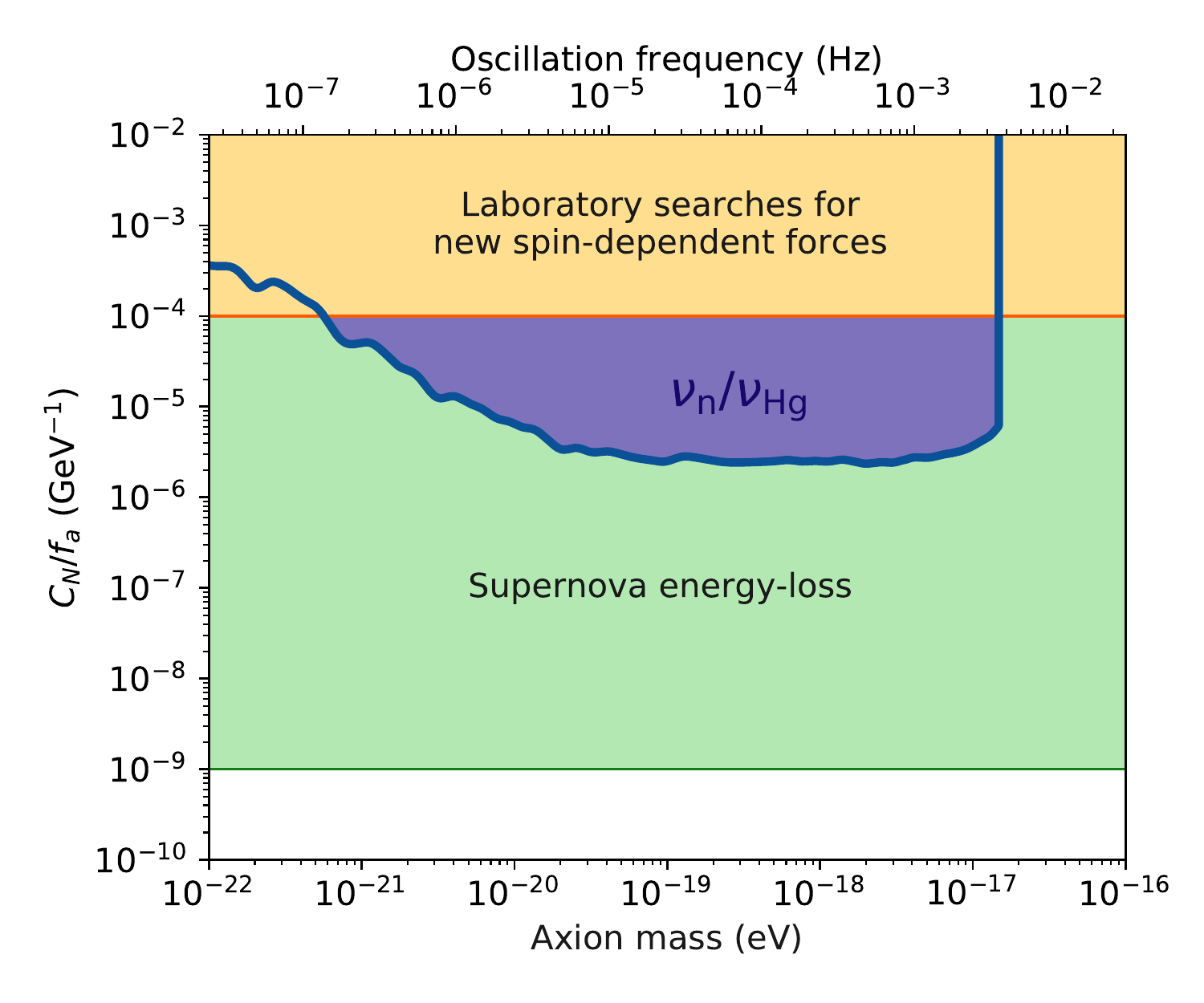}
  \caption{
Limits on the interactions of an axion with the gluons (top) and nucleons (bottom), as defined in Eq.\,(\ref{Axion_couplings}), assuming that axions saturate the local cold DM content.
  The regions above the thick blue and red lines correspond to the regions of parameters excluded by the present work at the 95\% confidence level (C.L.).
  The colored regions represent constraints from Big Bang nucleosynthesis (red, 95\% C.L.) \cite{Blum2014,StadnikThesis,Stadnik2015D}, supernova energy-loss bounds (green, order of magnitude) \cite{Graham2013,Raffelt1990Review,Raffelt2008LNP}, consistency with observations of galaxies
  %(orange, \note{@DM: What is the C.L. here?})
  (orange) \cite{Marsh2015Review,Marsh2015B,Schive2015,Marsh2017}, and laboratory searches for new spin-dependent forces (yellow, 95\% C.L.) \cite{Romalis2009_NF}.
  %\note{DM: give approximate C.L. or refer to text for explanation of level of rigour}, and consistency with observations of galaxies (orange) \cite{Marsh2015Review,Marsh2015B,Schive2015,Marsh2017}.
  The nEDM, {\color{black}$\nu_\mathrm{n} / \nu_\textrm{Hg}$} and Big Bang nucleosynthesis constraints scale as $\propto \sqrt{\rho_a}$, while the constraints from supernovae {\color{black}and laboratory searches for new spin-dependent forces} are independent of $\rho_a$.
  The constraints from galaxies are relaxed if axions constitute a sub-dominant fraction of DM.
  We also show the projected reach of the proposed CASPEr experiment (dotted black line)~\cite{CASPEr2014}, and the parameter space for the canonical QCD axion (purple band).
%   \note{DM: y-axis should be $\sqrt{\rho}$}
%   \note{YS: Actually, I suggest to remove $\rho_a$ altogether from the y-axis label, since the QCD axion and Super-Planckian regions are $\rho_a$-indepedent, as too are the supernova energy-loss bounds. The scaling of the constraints is in fact already explained in the caption above: we state that the constraints assume saturation of DM by axions (which is correct for all of the regions on the graph), then we proceed to state that the nEDM and BBN constraints scale as $\propto{\sqrt{\rho_a}}$, so that the reader will have in mind the $\propto{\sqrt{\rho_a}}$ implicitly on the y-axis for these constraints.}\note{DM: okay, but if we assume saturation of DM then some (even dotted or mild line) for galaxies should go in?}
%\note{KK: a comment on 'Super-Planckian' is missing.}
%\note{YS: I have now added a brief comment on this in the main text.}
% \note{PH: our lines need to go up to the top of the plot.  There has to be a division between excluded and not excluded regions.}
% \note{YS: (Depending on the changes proposed above,) for aesthetics, could we perhaps shift the "Super-Planckian axion decay constant" label downwards? And perhaps also shift the "long time-base" label down and to the left, and shift the "short time-base" label down and a bit to the left?}
}
\label{fig:axion_limits_v2}
\end{figure}

We also perform a search for the axion-wind effect, Eq.\,(\ref{potential_axion-wind}), by partitioning the entire PSI dataset into two sets with opposite magnetic-field orientations (irrespective of the electric field) and then analyzing the ratio $R = \nu_\mathrm{n} / \nu_\textrm{Hg}$ similarly to our oscillating EDM analysis above.
The axion-wind effect would manifest itself through time-dependent shifts in $\nu_\mathrm{n}$ and $\nu_\textrm{Hg}$ (and hence $R$) at three angular frequencies:~$\omega_1 = m_a$, $\omega_2 = m_a + \Omega_{\textrm{sid}}$ and $\omega_3 = |m_a - \Omega_{\textrm{sid}}|$, with the majority of power concentrated in the $\omega_1$ mode.
Also, the axion-wind signal would have an opposite phase in the two subsets.
We find two overlapping $3\sigma$ excesses in the two subsets (at $\unit[3.42969]{\upmu Hz}$ and $\unit[3.32568]{mHz}$), neither of which have a phase relation consistent with an axion-wind signal.
Following Eq.\,(\ref{potential_axion-wind}), we derive limits on the axion-nucleon coupling in Eq.\,(\ref{Axion_couplings}).
We present these limits in Fig.\,\ref{fig:axion_limits_v2}, assuming that axions saturate the local cold DM energy density.
Our peak sensitivity is
$f_a/C_N \approx \unit[4 \times 10^{5}]{GeV}$ for $\unit[10^{-19}]{eV} \lesssim m_a \lesssim \unit[10^{-17}]{eV}$.

\section{Conclusions}
In summary, we have performed a search for a time-oscillating neutron EDM in order to probe the interaction of axion-like dark matter with gluons.
We have also performed a search for an axion-wind spin-precession effect in order to probe the interaction of axion-like dark matter with nucleons.
So far, no significant oscillations have been detected, allowing us to place limits on the strengths of such interactions.
Our limits improve upon existing astrophysical limits on the axion-gluon coupling by up to 3 orders of magnitude and also improve upon existing laboratory limits on the axion-nucleon coupling by up to a factor of 40.
Furthermore, we constrain a region of axion masses that is complementary to proposed ``on-resonance'' experiments in ferroelectrics \cite{CASPEr2014}. Future EDM measurements will allow us to probe even feebler oscillations and for longer periods of oscillation that correspond to smaller axion masses.

%\note{Rough checklist: --------------------------------------------}
%\note{Descriptions of experimental procedures/methods employed in ILL and PSI experiments}
%\note{Descriptions of analysis methods employed, including discussion of criteria used to distinguish real axion DM signal from false signals}

%\note{1 sample periodogram of ILL data and 1 sample periodogram of PSI data, as a minimum}

%\note{Graph showing limit on oscillating neutron EDM (strictly, the weighted linear combination with mercury EDM), as a function of frequency}

%\note{Graph showing our constraints on axion-gluon coupling, for the axion mass range $10^{-24}~\textrm{eV} \le m_a \le 10^{-17}~\textrm{eV}$.
%Include BBN limits for comparison.
%Is it also worth showing the supernovae limits here (which are much weaker))? }

%\note{In the ensuing discussion of our results, mention that we probe a complementary range of axion masses to the proposed CASPEr-Electric experiment \cite{CASPEr2014}.
%}

%%%%%%%%
%\section*{ACKNOWLEDGEMENTS}
\begin{acknowledgments}
We are grateful to Maxim Pospelov for helpful discussions.
The experimental data has been taken in part at the ILL Grenoble and at PSI Villigen. We acknowledge the excellent support by the technical groups of both institutions and by various services of the collaborating universities and research laboratories. Dedicated technical support by M.~Meier and F.~Burri is gratefully acknowledged.
We remember with gratitude the pioneering contributions of Professors K.~Smith and J.~M.~Pendlebury, without whom these experiments could never have taken place. This work was funded in part by the U.~K.~Science and Technology Facilities Council (STFC) through grants ST/N000307/1 and ST/M503836/1, as well as by the School of Mathematical and Physical Sciences at the University of Sussex.   The original apparatus at ILL was funded by grants from the U.~K.’s PPARC (now STFC), and we would like to thank the generations of engineers, students and Research Fellows who contributed to its development.
We gratefully acknowledge support of the Swiss National Science foundation under grants number 200020\_172639, 200020\_163413, and 200020\_157079.
% SNF grants 200020_163413, 200020_157079
%Support by the Swiss National Science foundation under grant number 200020\_172639 (MR, JK at ETHZ) is gratefully acknowledged.
This work has been supported in part by The National Science Centre, Poland, under the grant No. UMO2015/18/M/ST2/00056.
This work has been supported by the Research Foundation - Flanders (FWO).
The LPC Caen and the LPSC acknowledge the support of the French Agence Nationale de la Recherche under Reference
No.~ANR-09-BLAN-0046.
M.~F.~was supported partly by the STFC Grant ST/L000326/1 and also by the European Research Council under the European Union's Horizon 2020 programme (ERC project 648680 DARKHORIZONS).
V.~V.~F.~was supported by the Gutenberg Research College Fellowship and by the Australian Research Council.
D.~J.~E.~M.~was supported by a Royal Astronomical Society postdoctoral fellowship hosted at King's College London.
P.~M.~M.~was supported by the State Secretariat for Education, Research and Innovation (SERI) - Federal Commission for Scholarships for Foreign Students (FCS) grant \#2015.0594.
%PMM: Additionally I wanted to bring to your notice that I would like to acknowledge the support from: SERI (State Secretariat for Education, Research and Innovation) - FCS (Federal Commission for Scholarships for Foreign Students) grant \#2015.0594.
Y.~V.~S.~was supported by the Humboldt Research Fellowship and in part by the Australian Research Council.
E.~W.~was supported by a PhD Fellowship of the Research Foundation - Flanders (FWO).
%EW: I am bound by my FWO contract to mention that I am a “PhD Fellow of the Research Foundation - Flanders (FWO)”. Could you mention this sentence somewhere in the acknowledgements?
\end{acknowledgments}

\bibliography{bib}

\begin{thebibliography}{89}
\bibitem [{tun()}]{tuningfootnote}%
  \BibitemOpen
  \href@noop {} {}\bibinfo {note} {Interaction (\ref{Axion_couplings}) also
  non-perturbatively induces a mass $m_a \approx 6C_G\,\mu\text{eV} \cdot
  (10^{12}\text{ GeV}/f_a)$. Axions with masses much smaller than this are
  theoretically fine-tuned.}\BibitemShut {Stop}%
\end{thebibliography}

\end{document}